\documentclass[twocolumn,pra,showpacs,multicol,amssymb]{revtex4}
\usepackage{graphicx}
\usepackage{dcolumn}
\usepackage{bm}
\usepackage{graphics}
\usepackage{epsfig,color}
\usepackage{tabu}
\usepackage{array}
\usepackage[none]{hyphenat} 

\newcommand{\be}{\begin{equation}}
\newcommand{\ee}{\end{equation}}
\newcommand{\bea}{\begin{eqnarray}}
\newcommand{\eea}{\end{eqnarray}}

\emergencystretch=\maxdimen
\begin{document}
\title{ Fidelity approach: the misleading role of finite size level crossing of excited energies   

}
\author{Somayyeh Nemati$^{1, 2, 3}$}
\author{Fatemeh Khastehdel Fumani$^{4}$} 
\author{Saeed Mahdavifar$^{4}$}

\affiliation{$^{1}$ Beijing Computational Science Research Center, Beijing 100193, China}
\affiliation{$^{2}$ Center of Physics of University of Minho and University of Porto, P-4169-007 Oporto, Portugal}
\affiliation{$^{3}$ University of Aveiro, Department of Physics, P-3810193 Aveiro, Portugal}
\affiliation{$^{4}$ Department of Physics, University of Guilan, 41335-1914, Rasht, Iran}

\begin{abstract}
Here, we show that, although quantum fidelity can truly identify two quantum phase transitions of a one-dimensional spin-1/2 quantum Ising model with competing nearest and next-nearest neighbor interactions in a transverse magnetic field, it may not be a suitable approach for analyzing its ground-state phase diagram.
\end{abstract}
\maketitle

We wish to point out misleading results that Bonfim \textit{et al.} obtain in their paper titled "Quantum fidelity approach to the ground-state properties of the one-dimensional axial
next-nearest-neighbor Ising model in a transverse field" \cite{Bonfim17} with the Hamiltonian
\begin{eqnarray}
{\cal H} &=& \sum_{n=1}^{N} (- J_{1}{\sigma}^{z}_{n}
{\sigma}^{z}_{n+1}+
 J_{2}{\sigma}^{z}_{n}
{\sigma}^{z}_{n+2})-B_x\sum_{n=1}^{N} {\sigma}^{x}_{n},\nonumber \\  
\label{H1}
\end{eqnarray}
where $\sigma_{n}$ shows the usual Pauli operator at the $n$-th site and $B_x$ denotes the transverse magnetic field. The interaction strength between nearest and next-nearest neighbors are both non-negative values and related to each other by $J_2=\alpha J_1$ which $\alpha (>0)$ is called frustration parameter. They numerically study the fidelity susceptibility as a function of the transverse field and the frustration parameter to determine the transition lines in the ANNNI ground-state (GS) phase diagram. According to their results, they claim that there are infinite numbers of modulated phases as well as ferromagnetic, floating, and $\langle 2,2 \rangle$ phases in the thermodynamic limit. Furthermore, their results show that the region with the floating phase similar to the modulated phases depends on the size of the system and will become considerably small in the thermodynamic limit.
\par
In recent decades, fidelity susceptibility as a concept of quantum information theory has been known as one of the candidates for identifying the quantum phase transition points \cite{You11, Zanardi06, Wang09}. It becomes maximum or sometimes diverges at the critical point \cite{Venuti07}. Its concept stems from quantum fidelity which basically is the overlap between two neighboring GSs of the system. Therefore, it is obviously a state-dependent parameter like the fidelity itself \cite{Jozsa94, You07}. The dependency of the fidelity susceptibility on the states makes a serious limitation for its application in determining quantum criticality of systems with degenerate states.

 The one-dimensional ANNNI model is originally a frustrated system and like most of the frustrated systems, its low energy states are challenging. For example, its GS degenerates at some certain coupling constants and transverse magnetic field values \cite{Jongh12}. 
In addition, low-lying excited-state (ES) energy level crossings happen in the ANNNI chain. Consequently, they may induce a serious problem that prevents reliable results in identifying the ANNNI phase boundaries when the fidelity susceptibility approach is applied. 
This is the crucial point that has not been taken into account in the Ref.~\cite{Bonfim17}.
 
\begin{figure*}[t!]
\centerline{\includegraphics[width=0.25\textwidth]{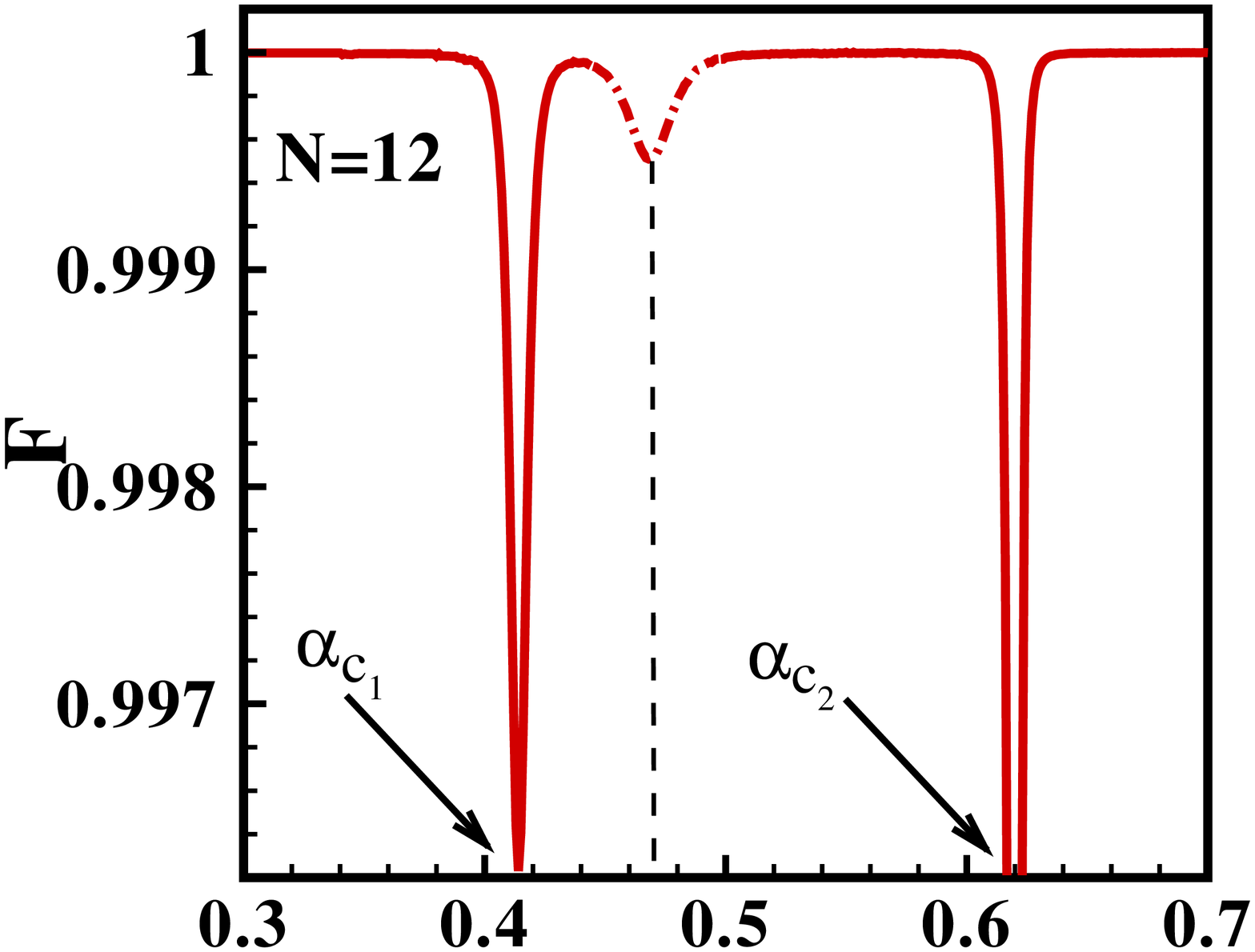}
\includegraphics[width=0.25\textwidth]{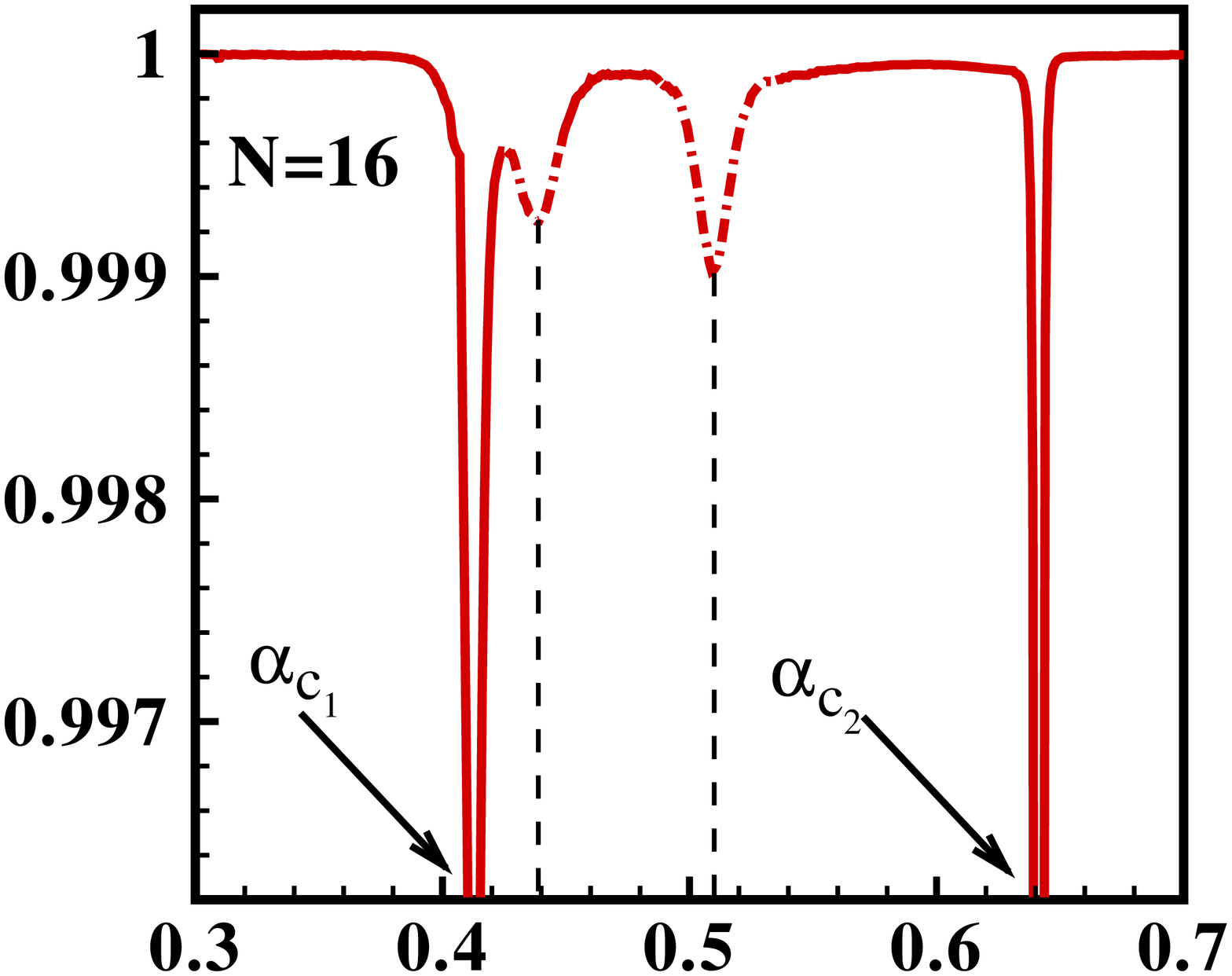}
\includegraphics[width=0.25\textwidth]{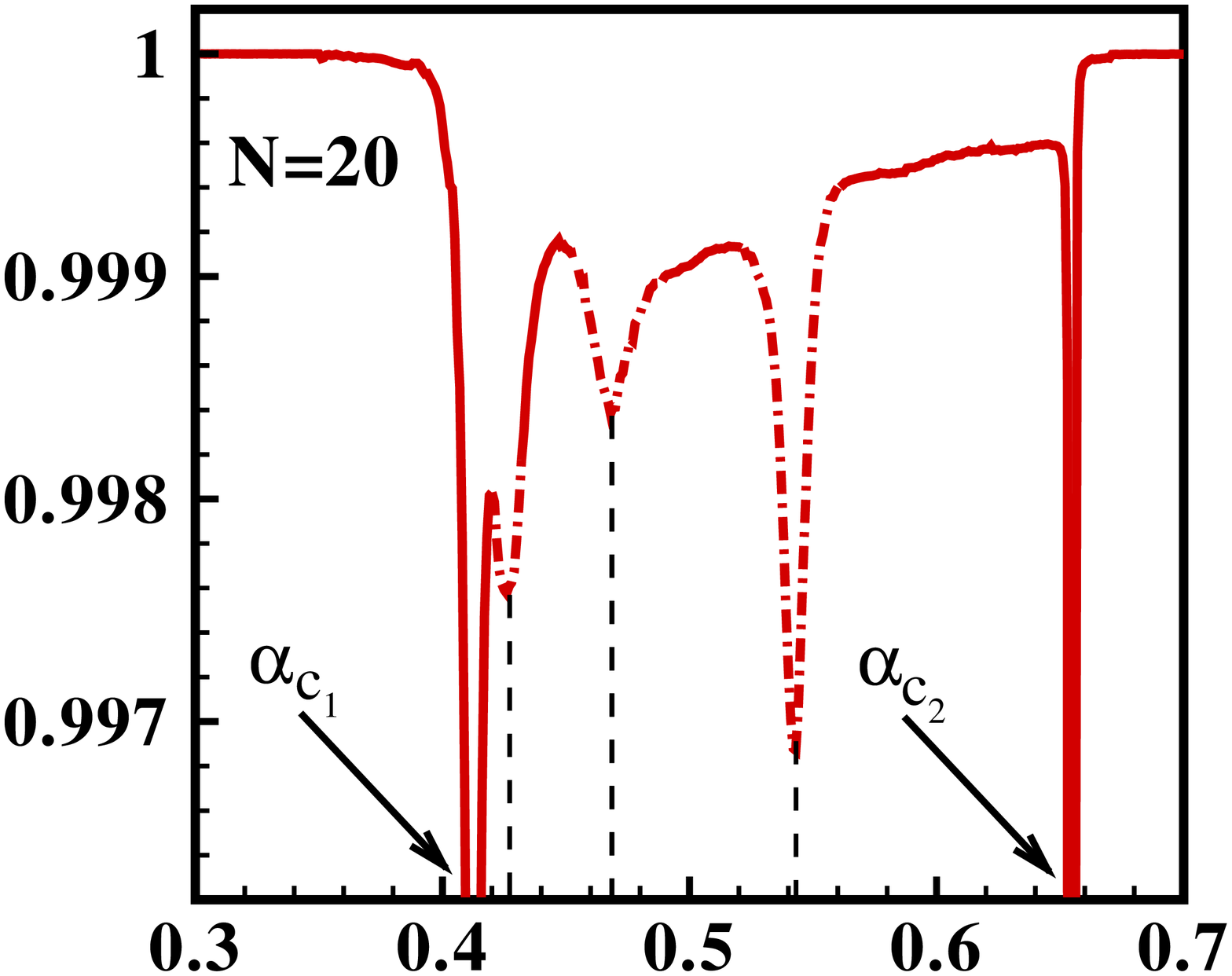}}
\centerline{\includegraphics[width=0.256\textwidth]{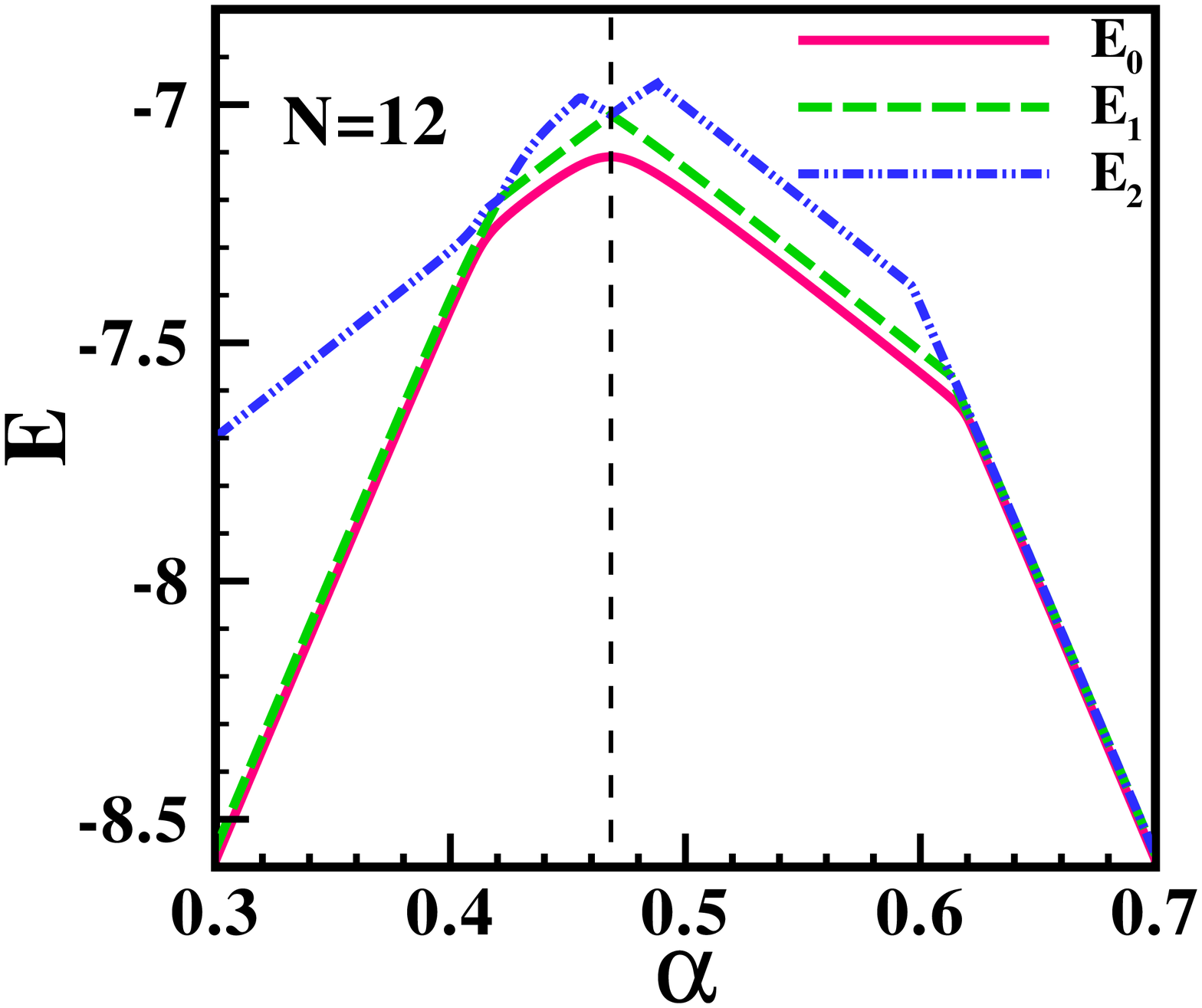}
\includegraphics[width=0.256\textwidth]{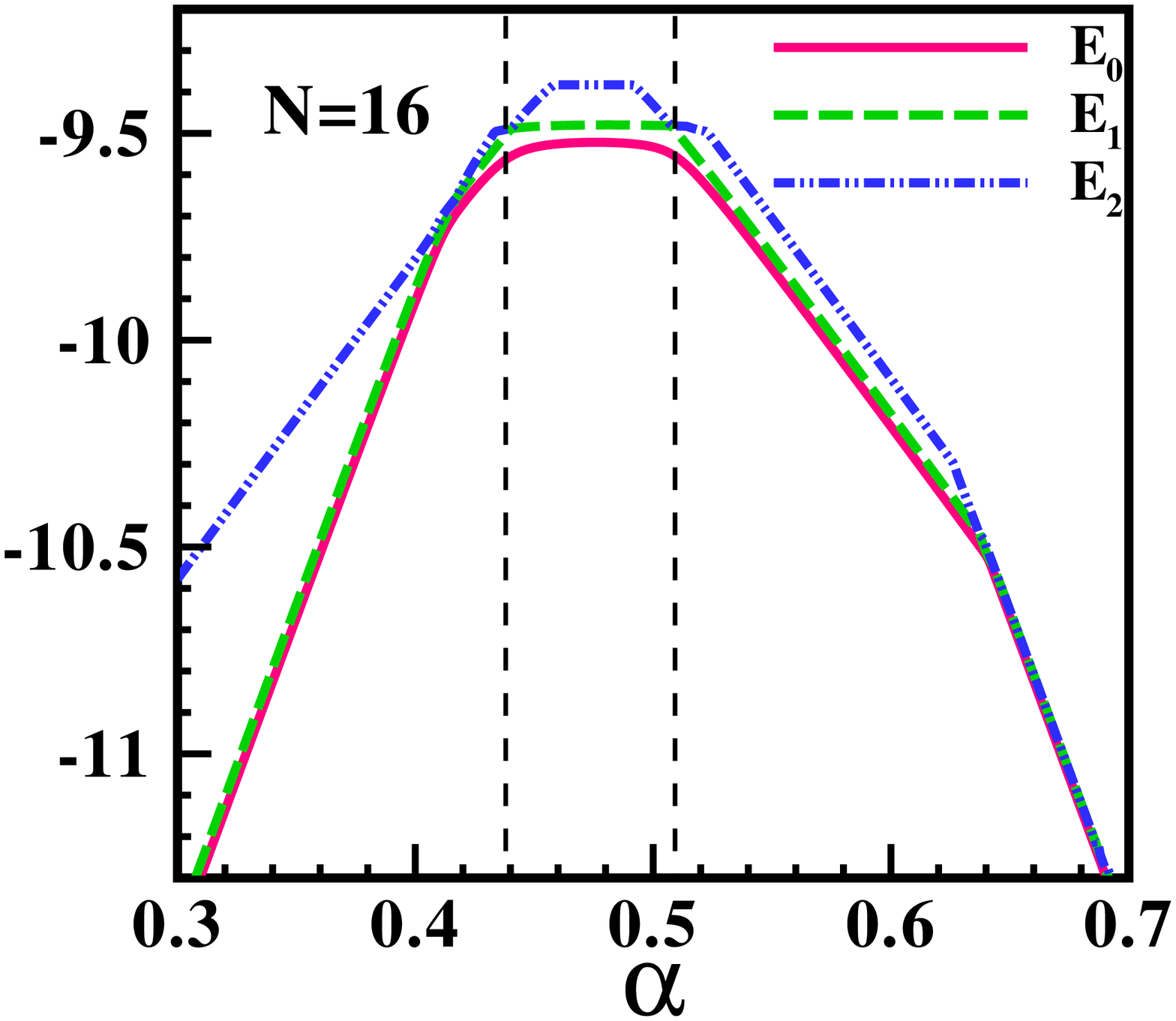}
\includegraphics[width=0.256\textwidth]{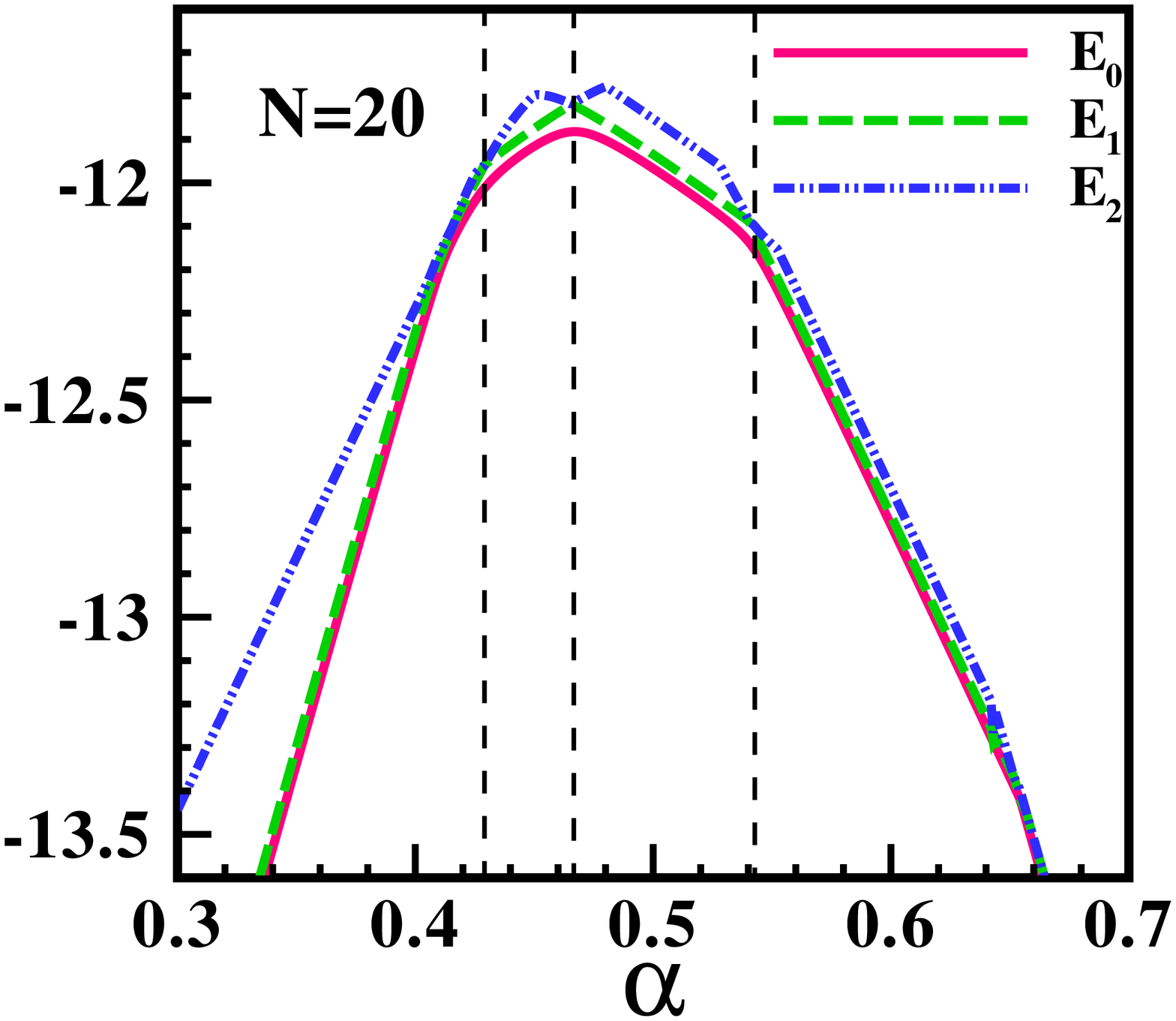}}
\caption{(Color online) Top panels left to right show the fidelity with respect to the frustration parameter for finite chain lengths $N=12, 16, 20$, respectively. Size-dependent drops of the fidelity are depicted by the dash-dotted lines. Two other drops of fidelity identify the quantum phase transition points shown by $\alpha_{c_1}$ and $\alpha_{c_2}$, respectively. Bottom panels left to right are diagrams of three lowest energy eigenvalues versus $\alpha$ for chains with $N=12, 16, 20$ spins, respectively. Vertical dashed lines in both top and bottom diagrams schematically denote one-to-one correspondence between the place of size-dependent drops of fidelity and the first ES energy level crossings in each considered chain size. In all diagrams the magnetic field has the fixed value as $B_x = 0.2$.}
\label{fig1}
\end{figure*}
\begin{figure}[b!]
\centerline{\includegraphics[width=0.25\textwidth]{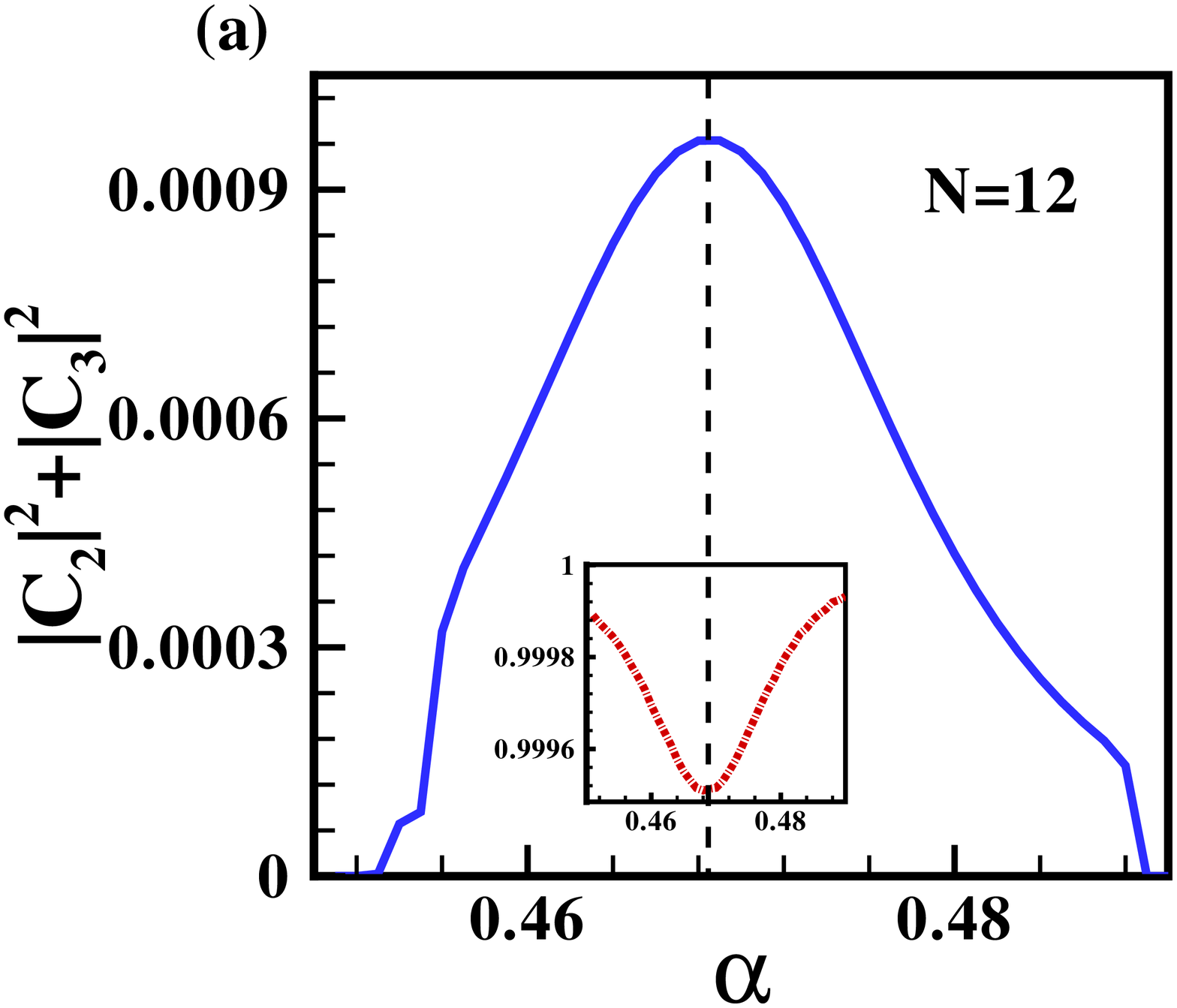}}
\centerline{\includegraphics[width=0.25\textwidth]{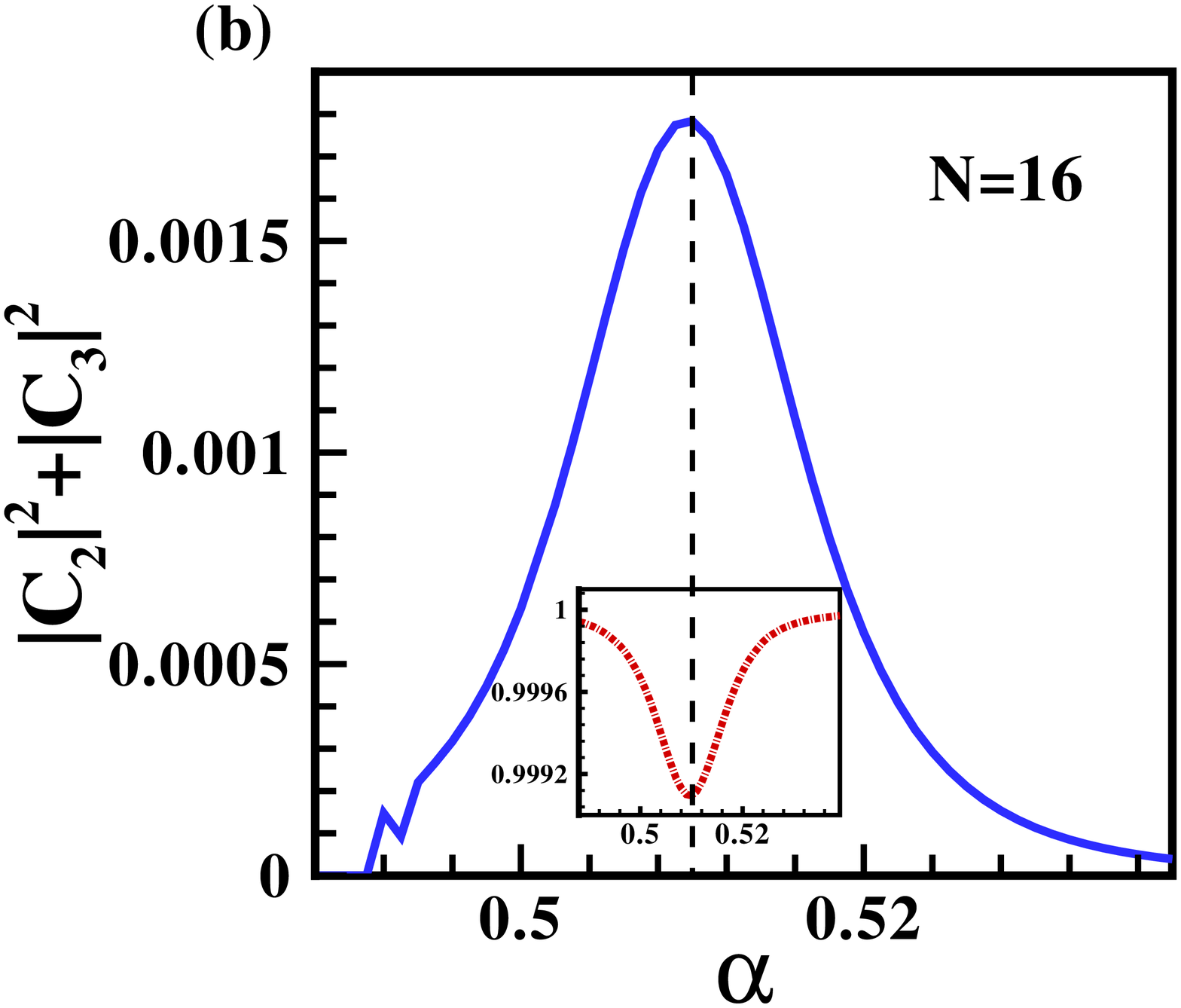}}
\caption{(Color online) The summation of $\mid C_2 \mid ^2$ and $\mid C_3 \mid ^2$, as a function of the frustration parameter, $\alpha$, for an ANNNI chain with a length of (a) $N=12$ and (b) $N=16$, respectively. The fidelity diagram appeared in the insets of both diagrams show complete coincidence between the maximum of this summation and the minimum of $F$ at $\alpha\sim 0.47$ and $\alpha\sim 0.51$ for $N=12$ and $N=16$ respectively.
} 
\label{fig2}
\end{figure}
\par
The result of the present comment provides strong evidence that predictions of Ref.~\cite{Bonfim17} for the GS phase diagram are not correct. We start with calculating the fidelity for the ANNNI chain and reach a conclusion that clearly justifies the invalidity of the fidelity approach in quantum critical points detection in the ANNNI chain.
\par
The fidelity is defined as the modulus of the overlap of normalized GS wave functions $| \Psi(\alpha) \rangle $ and $| \Psi(\alpha+\delta \alpha) \rangle $ for closely spaced frustration parameter $\alpha$ and  $\alpha+\delta \alpha$ as 
\begin{eqnarray}
F=| \langle \Psi(\alpha) | \Psi(\alpha+\delta \alpha) \rangle |.
\label{F}
\end{eqnarray}
Let us assume that the Hamiltonian $H(\alpha+\delta \alpha)$ has  eigenvectors $| m \rangle $ and eigenvalues $E_m$, so that $H(\alpha+\delta \alpha) | m \rangle = E_m | m \rangle$. Writing the GS $|\Psi(\alpha) \rangle$ as $| \Psi(\alpha)\rangle=\sum_{m} C_m | m \rangle$, the fidelity will become $F=\sqrt{1-\sum_{m \neq 0} |C_m|^2}$, where $C_m$ is the probability amplitude of finding the system in $m-$th ES of the Hamiltonian $H(\alpha+\delta \alpha)$.

Using the numerical Lanczos method we have diagonalized the Hamiltonian for finite-size chains and the GS and low energy eigenvectors are obtained. In Fig.~\ref{fig1}, we have plotted the fidelity and the three lowest energy eigenvalues of the system as a function of the frustration parameter, $\alpha$, for the chains with size $N=12,16,20$ in a transverse magnetic field $B_x=0.2$. As it can be seen, the fidelity drops at the points where either the GS or first ES finite-size level crossings happen since the probability amplitudes of finding the system in low-lying ESs of the Hamiltonian $H(\alpha+\delta \alpha)$ can have considerable values at these points.
 It is very important to note that the mentioned phenomenon is rooted in the fact that $H(\alpha)$ does not commute with $H(\alpha+\delta \alpha)$. Later, we explicitly show that the first ES level crossings play a misleading role in fidelity behavior.
 
In principle, at both kinds of level crossing points, the GS or first ES energies become at least two-times degenerate.
 As is clear, more first ES level crossings (and subsequently more dropped points in the fidelity) happen in larger ANNNI chains. Notice that only the GS level crossing point can characterize a phase transition. The location of two GS level crossing points that appear in all considered ANNNI chains in the energy diagrams of Fig. \ref{fig1} do not significantly vary with the changing of the system size. As the authors of the original paper are correct in identifying the ferromagnetic-paramagnetic and $\langle 2,2 \rangle$-floating transition points  at $\alpha_{c_1}\sim 0.42$ and $\alpha_{c_2}\sim 0.64$ respectively, the mentioned GS level crossings can truly denote both of them. 
 
 We only concentrate on the region of frustration parameter with size-dependent first ES level crossings and fidelity minimum points in the considered system. To this aim, we study another appropriate parameter as a summation of complex square of the ESs probability amplitudes, $\sum_{m \neq 0} \mid C_m\mid^2$ that can clarify the fidelity behavior. Numerically, we have checked up to the five first ES eigenvectors of $H(\alpha+\delta\alpha)$ and calculated their probability amplitudes ($C_m$). Our analysis clarified that $|C_2|^2$ and $|C_3|^2$ have the main role in the $\sum _{m\neq 0}|C_m|^2$ at first ES level crossing points. Fig.~\ref{fig2} (a) and (b) show $\mid C_2\mid^2+\mid C_3\mid^2$ versus the $\alpha$ for an ANNNI chain with $N=12$ and $N=16$ spins, respectively, in the vicinity of a considered first ES level crossing point. Interestingly, it becomes maximum at the points where the first ES level crossing happens.
 It is notable that these local peaks change with the size of the system and consequently cause minimum points in the fidelity behavior as is seen in the insets of Fig.~\ref{fig2} (a) and (b). 

\par
To sum up, our results reveal that all minimum points of fidelity that appear as the maximum points of fidelity susceptibility cannot be the sign of critical points in the ANNNI chain. Using fidelity and functions driven by fidelity are not always appropriate for the study of the quantum phase transitions. It should be checked and confirmed that the results of fidelity susceptibility are valid for the system of interest.


\end{document}